\documentclass[conference]{IEEEtran}

\usepackage[colorlinks=true, linkcolor=blue, citecolor=blue, urlcolor=blue]{hyperref}

\usepackage{amsmath}
\usepackage{balance}
\usepackage{graphicx}
\usepackage{url}

\usepackage{color}
\newcommand{\change}[1]{\textcolor{black}{#1}}

\newcommand{\researchQuestion}{\textmd{RQ}}
\newcommand{\researchQuestionConstraints}%
{\researchQuestion1}
\newcommand{\researchQuestionRequirements}%
{\researchQuestion2}
\newcommand{\researchQuestionArchitectures}
{\researchQuestion3}

\newcommand{\constraint}{\textmd{C}}
\newcommand{\constraintCustomerHeterogenity}%
{\constraint1}
\newcommand{\constraintTechnologicalDependency}%
{\constraint2}
\newcommand{\constraintSizeOfProject}%
{\constraint3}
\newcommand{\constraintExternalInvestments}%
{\constraint4}
\newcommand{\constraintVarietyOfServers}%
{\constraint5}
\newcommand{\constraintUpdateOfOldSystems}%
{\constraint6}
\newcommand{\constraintReusability}%
{\constraint7}
\newcommand{\constraintMaintenanceAndSupport}%
{\constraint8}
\newcommand{\constraintDistributionOfPersonalData}%
{\constraint9}

\newcommand{\requirement}{\textmd{R}}
\newcommand{\requirementSystemSecurityAndPrivacy}%
{\requirement1}
\newcommand{\requirementDataMinimization}%
{\requirement2}
\newcommand{\requirementConstentControl}%
{\requirement3}
\newcommand{\requirementDataTraceability}%
{\requirement4}
\newcommand{\requirementUserAccess}%
{\requirement5}
\newcommand{\requirementDataRectification}%
{\requirement6}
\newcommand{\requirementDataErasure}%
{\requirement7}
\newcommand{\requirementDataRestrictions}%
{\requirement8}
\newcommand{\requirementPhysicalLocationOfData}%
{\requirement9}

\begin{document}

\title{The General Data Protection Regulation: Requirements, Architectures, and Constraints}
\author{
\IEEEauthorblockN{Kalle Hjerppe}
\IEEEauthorblockA{Geniem Oy, Finland \\
Email: kalle.hjerppe@gmail.com}
\and
\IEEEauthorblockN{Jukka Ruohonen}
\IEEEauthorblockA{University of Turku, Finland \\
Email: juanruo@utu.fi}
\and
\IEEEauthorblockN{Ville Lepp\"anen}
\IEEEauthorblockA{University of Turku, Finland \\ 
Email: ville.leppanen@utu.fi}
}

\maketitle

\begin{abstract}
The General Data Protection Regulation (GDPR) in the European Union is the most famous recently enacted privacy regulation. Despite of the regulation's legal, political, and technological ramifications, relatively little research has been carried out for better understanding the GDPR's practical implications for requirements engineering and software architectures. Building on a grounded theory approach with close ties to the Finnish software industry, this paper contributes to the sealing of this gap in previous research. Three questions are asked and answered in the context of software development organizations. First, the paper elaborates nine practical constraints under which \change{many} small and medium-sized enterprises (SMEs) \change{often} operate when implementing solutions that address the new regulatory demands. Second, the paper elicits nine regulatory requirements from the GDPR for software architectures. Third, the paper presents an implementation for a software architecture that complies both with the requirements elicited and the constraints elaborated.
\end{abstract}


\section{Introduction}

The famous GDPR became enforceable in the European Union (EU) in late May 2018 ~\cite{EU16a}.\footnote{~This paper is based on the M.Sc.~thesis of Kalle Hjerppe.} The regulation's political preparations as well as its later legal enforcement were received with both great enthusiasm and great anxiety. \change{Yet}, a~familiar theme during the past two years or so was the ill-preparation of many companies to live in a post-GDPR world. Even in Finland, where data protection regulation already had a relatively strong legislative ground, many companies were either unaware of the GDPR's implications or reluctant to implement the changes required~\cite{Mikkonen14}. The same theme has continued even after \change{over} a year of the GDPR's enforcement. While questionable, \change{unethical, and possibly illegal} data collection practices may explain some of the background behind the continuance, another explanation culminates to the fact that \change{only little technical guidance has been provided by the EU} for implementing the changes required. Even the GDPR's requirements are still somewhat unclear and much debated.

\change{These points extend to academic research. Although some previous work has been done on eliciting requirements from the GDPR~\cite{AyalaRivera18, Ringmann18}, the understanding of the regulation's full scope is still in its infancy.} Thus, there is a clear need for a hands-on demonstration on how the GDPR's regulatory demands are implemented in the contemporary software industry. Such a demonstration cannot rely on requirements engineering alone; something must be said also about the context within which day-to-day software engineering operates, and something concrete must be shown in order to demonstrate the plausibility of the requirements elicited. \change{These practical aspects are particularly important for SMEs~\cite{Brodin19}. While large companies have vast resources for maintaining in-house legal departments, using consultants, and hiring dedicated privacy engineers, most smaller companies operate with tight budgets and other related constraints that influence also the technical implementation of data protection and privacy solutions.}

\change{Given this practical motivation, the paper presents a case study on implementing requirements elicited from the GDPR in a Finnish SME. In addition to the framing to SMEs and the Finnish software industry, the paper's scope is restricted to software architectures in general and so-called \textit{service-oriented architectures} (SOAs) in particular. With this framing,} the following three \textit{research questions} (RQs) are examined:

\begin{itemize}
\item{\researchQuestionConstraints: \textit{What practical constraints SMEs \change{operating with SOAs} typically face when implementing solutions for complying with the GDPR's new regulatory demands?}}
\item{\researchQuestionRequirements: \textit{What requirements does the GDPR imply \change{for SOAs operated by SMEs,} and how these can be elicited?}}
\item{\researchQuestionArchitectures: \textit{How the requirements elicited from the GDPR can be reasonably implemented \change{in SOAs} under the practical constraints that many \change{software development} SMEs face?}}
\end{itemize}

The structure of the paper's remainder follows these three research questions. After the opening Section~\ref{sec: background} that briefly outlines the background and related work, Section~\ref{sec: context} elaborates the context, \change{methods, and} the constraints identified, Section~\ref{sec: requirements} presents the requirements elicitation, and Section~\ref{sec: architecture} illustrates the \change{main changes to the company's} software architectures. The final Section~\ref{sec: discussion} summarizes the answers to the research questions, notes \change{limitations} and a few practical challenges faced, and pinpoints directions for further research.

\section{Background and Related Work}\label{sec: background}

Requirements engineering first appeared in the software engineering literature in the early 1980s. It was already back then framed with the classical dual questions about software \textit{validation} (``building the right product'') and the \textit{verification} (``building the product right'') of the software produced~\cite{Boehm84}. In a traditional software engineering setup the former question connotes with an elicitation of requirements from stakeholders. The latter question includes assertions about whether the requirements elicited match the implementation produced, and further whether the implementation meets quality attributes deemed as relevant. Both questions are notoriously difficult to answer in practice. In fact, requirements engineering is among the fundamental factors in the complex causal constellation through which software projects fail~\cite{Lehtinen14}. \change{Furthermore,} software engineering is increasingly facing also various difficult societal requirements~\cite{Ruhe17}. Software engineering should innovate but still stay ethical and sometimes even conservative; software should encourage and empower at the same time as it should curtain the negative effects from these; software should be safe and secure but still \change{usable and} easily accessible; and so forth.

Legal regulations are the primary force through which these new requirements enter into the realm of software engineering. Information security is one thing. Although there has long been some regulatory requirements about (information) security, many of these have been limited to some particular software industry sectors. Privacy is another thing. When compared to security, privacy and data protection regulations are relatively new---the GDPR is the first regulation covering a large geographic region. Even though there is still a lot of room for improvements, there have long been also many frameworks, tools, guidelines, and checklists for meeting different security requirements~\cite{Elluri18, Rindell18}. In some cases, even formal audits and certification are possible. In contrast, very little guidance exists for addressing requirements from the GDPR and other emerging regulations for privacy and data protection.

Notions such as ``privacy-by-design'', ``privacy-by-default'', and ``data-protection-by-design'' frequently appear in the legal literature~\cite{Mikkonen14, TikkinenPiri18, Veale18}, but the work has just recently begun on how the required paradigm change could be achieved in software engineering and computer science in general. While there are some good practical examples, including those related to distributed software architectures~\cite{Kittman18, SuHyysalo16}, there exists also some empirical evidence that software engineers are even actively discouraged from making privacy a priority~\cite{Hadar18}. Web privacy is a good example about the latter point. In this domain the GDPR's impacts have so far been debatable at best, and the potential explanations include a lack of practical guidance and a false sense of compliance~\cite{Degeling19}. By and large, these basic explanations presumably generalize also to other domains and software industry sectors. In order to improve the situation, requirements engineering should be in a focal point. After all: when it is unclear whether the right product is being built, it is also unlikely that the product is being built right. 

Requirements engineering contains numerous frameworks, tools, and guidelines for eliciting requirements from stakeholders. \change{However, these are largely just emerging for eliciting requirements from regulations instead of stakeholders.} A basic problem with regulations is that the question is not only about elicitation of requirements from the regulations but also about legal compliance to the regulations~\cite{Olislaegers11}. By implication, the elicitation should be especially rigorous and the requirements particularly sharp. In theory, a sensible approach might involve both lawyers and software engineers, but even this collaborative approach contains plenty of problems. For instance, both lawyers and engineers speak their own more or less formal languages but with vastly different dialects; both typically have only a limited understanding about each other's domains and knowledge; and so forth~\cite{Bobkowska10}. Even when such obstacles could be remedied, it should be emphasized that small software development enterprises seldom have the resources to collaborate with lawyers during requirements elicitation---let alone in the whole development process during which stakeholders should be involved in agile software development. Further problems are caused by the fact that like all requirements, also requirements elicited from legal regulations change over time~\cite{Gordon13}. Stakeholders and users may also have non-regulatory security, privacy, and data protection requirements of their own \cite{Gharib16}. Given these and other subtle problems, how is it possible to address the GDPR in practice?

\change{Because} the fundamental question about legal requirements remains more or less unaddressed in the existing literature, a so-called \textit{argumentative requirements engineering}~\cite{Ingolfo13, Muthuri16} provides a sensible theoretical frame for the present work. Thus, in what follows, the solution proposed ($\researchQuestionArchitectures$) is argued to comply with the GDPR given a set of practical constraints ($\researchQuestionConstraints$) and requirements elicited from the regulation ($\researchQuestionRequirements$). Actual compliance is left unaddressed. \change{There are} two layers:

\begin{small}
\begin{quote}
``The cost to maintain this argumentation-based treatment of compliance though, is the introduction of additional elements in the models, which may result in comprehension bottlenecks. We address this problem by splitting the model in 2 layers: the first (\textit{argumentation}) layer models a discussion concerning a compliance solution; the second (\textit{solution}) layer models the actual compliance solution. The argumentation layer records the structure and other information about the discussions evaluating
changes of the solution layer.'' \cite[p.~281]{Ingolfo13}
\end{quote}
\end{small}

Here, the argumentative layer addresses the constraints and requirements. The solution layer operates with concrete software \change{architectures} based on the arguments put~forward. \change{This dual layering is generally important for underlining the intrinsic value of argumentative requirements engineering in the present context: if a company must legally defend its data protection and privacy practices, both layers are necessary.}

\section{Context}\label{sec: context}

\subsection{The Case Company}

The analysis presented builds on a case study about the design of GDPR-induced \change{changes} for software architectures in a particular company. The company itself is a fairly typical, already well-established SME in the Finnish software industry. The case company's business model relies on what might be called ``software-engineering-as-a-service'' model: the company implements software projects for customers. In essence, each project produces a service, which is deployed on behalf of a customer on the company's deployment infrastructure. Agile software development practices are used during implementation. Different software modules and frameworks are used extensively to increase flexibility and reusability. While each project typically has a fixed length, these may be sometimes further developed. Customers usually also outsource maintenance tasks to the company. Increasingly, the company is also using so-called DevOps practices to improve different delivery and maintenance tasks. All in all, the case company can be interpreted as a typical representative in its respective domain.

\subsection{\change{Approach}}

\change{The research methodology builds on the so-called \change{(constructivist)} \textit{grounded theory} approach~\change{\cite{ChenBabar13, Stol16}}.} In general, grounded theory approaches are valuable for understanding how the GDPR is implemented in practice~\cite{Altorbaq17}. The research methodology adheres also to the principles of so-called \textit{design science}~\change{\cite{Brodin19, Hevner04}}. \change{In other words, the goal was to design changes to concrete software architectures. These changes not only take} regulatory requirements into account, but also \change{consider} the larger organizational, infrastructural, technological, and business constraints. \change{By implicitly following the so-called \textit{twin peaks} model for requirements and architectures~\cite{ClelandHuang13},} the \change{architectural changes were thus} ``constructed'' through ``data'' from the GDPR as well as from prior knowledge within the case company---\change{and} within comparable companies in the Finnish software industry between 2014 and 2017. \change{Although no systematic framework was used for dealing with the secondary, context-specific material, the engineering of the architectural solutions was not unsystematic. The following five auxiliary ``data sources'' are worth briefly remarking.}

\change{First and foremost, the choices are based on knowledge about the existing software architectures. For implementing the changes to these, the architectures were systematically analyzed together with the old design documents and requirements that led to the architectures. Second, the pre-GDPR legislation in Finland was analyzed together with the Finnish governmental guidance for information and software security. This guidance (abbreviated as VAHTI; see~\cite{Rindell18}) was used for addressing some of the GDPR's security-specific requirements. Third, discussions were held with some existing customers, and an external law firm was consulted regarding a few non-technical data protection questions. Fourth, all concrete changes went through code review and the design choices through more general peer review within the case company. Last, the changes were thoroughly tested, as always. Given these information sources, all typical design evaluation methods~\cite[p.~86]{Hevner04} were used except experimental~methods.}

\subsection{Constraints}\label{subsec: constraints}

Both the company's internal software engineering practices and the external business factors place constraints for \change{implementing the requirements elicited from the GPDR's articles.} The following nine \textit{constraints}~($\constraint$) were identified prior to the requirements elicitation and architectural design phases:

\begin{itemize}
\item{\textit{The heterogeneity of customers} ($\constraintCustomerHeterogenity$) imposes some constraints because there are both new customers and old customers with their already implemented projects. Particularly new customers may bring additional requirements regarding the software frameworks used in a project. In terms of the GDPR, the important implication is that the software architecture and its data structures cannot be tied to a single database holding personal data.}
\item{\textit{A technological dependency} ($\constraintTechnologicalDependency$) constraint follows directly from Constraint~$\constraintCustomerHeterogenity$. In other words, the requirements of both new and old customers influences the technological choices made. As always, there is also a certain price to pay from a commitment to a particular technology, including any software framework adopted.}
\item{\textit{The size of projects} ($\constraintSizeOfProject$) carried out in the company brings a further constraint. These range from small one-month projects to large undertakings spanning several years with annual budgets of hundreds of thousands of euros. Although the largest projects are excluded from the present work, $\constraintSizeOfProject$ implies that the architectural design should scale at least to middle-range and small projects.}
\item{\textit{External investments} ($\constraintExternalInvestments$) must be also taken into account. While it might be possible to design and implement a perfect-fit architecture for a single customer, such an architecture unlikely scales to projects of other customers. From the company's perspective, it is thus beneficial to design and implement as much functionality as is possible into unified frameworks shared in all projects. Due to the company's business model, licensing, intellectual property, and related aspects are also a part of $\constraintExternalInvestments$, although these are excluded from the present work.}
\item{\textit{The variety of servers} ($\constraintVarietyOfServers$) is one of the infrastructural characteristics that is particularly important in terms of data protection~\cite{Shastri19}. The final services implemented use caching servers and load balancers. The application servers may be distributed. Databases are typically deployed on servers of their own. Furthermore, a particular emphasis should be placed on the so-called staging environment, which is used during development and testing to replicate an environment used for the final deployment. Consequently, personal data may scatter to multiple servers, and the staging environment should hold the same data as the deployment environment. These aspects impose important constraints that should be carefully addressed to meet the requirements from the GDPR.}
\item{\textit{A need to update older systems} ($\constraintUpdateOfOldSystems$) is also apparent because data protection has been considered with varying scrutiny in older projects and their architectures. While these conform with the older data protection legislation in Finland, it is beneficial for the company to unify all related functionality into a single umbrella solution. Obviously, the unification must operate with available resources; therefore, the present work does not consider those older systems that would have required a substantial redesign. For these legacy systems, the GDPR's requirements were addressed with cheaper customized patching.}
\item{\textit{Reusability} ($\constraintReusability$) is a classical software engineering goal. As already mentioned with respect to $\constraintExternalInvestments$ and $\constraintUpdateOfOldSystems$, achieving the goal is constrained by many practical aspects, however. At the software architecture level, reusability connotes with concepts such as modularity, which introduces also the dual concepts of cohesion and coupling. In terms of data protection, the famous single responsibility principle~\cite{Martin03} provides a sensible goal; personal data should be isolated into particular modules designed to handle this type of data. Analogously to the infrastructural constraints, such isolation must balance the need to modify and customize other software modules, however.}
\item{\textit{Maintenance and support} ($\constraintMaintenanceAndSupport$) aspects must be also taken into account. In addition to traditional maintenance questions (such as implementing bug fixes and patching online deployments), it is important to emphasize different support services provided to customers. In general, these imply a necessity to separate different realms and roles from each other~\cite{MoralGarcia14}. In other words, the architecture must conform with roles assigned to customers, developers, system administrators, and support personnel. By implication, $\constraintMaintenanceAndSupport$ includes not only access control mechanisms but also considerations related to logging.}
\item{\textit{Distribution of personal data} ($\constraintDistributionOfPersonalData$) is the last but not the least constraint. In addition to the distribution of personal data at the infrastructural level ($\constraintVarietyOfServers$), personal data may distribute across multiple places and levels of a software architecture. Multiple modules may handle such data. Also granularity must be addressed: multiple implementation units (such as classes and configuration files) may deal with sensitive personal data. Given this background, the design followed a simple but fundamental principle: the lesser the distribution of personal data, the better the solution for data protection. This principle connotes with those more generally related to information security. The principle can be further seen as a software architecture equivalent to the GDPR's data minimization principle.}
\end{itemize}

\subsection{An Example Architecture}\label{subsec: an example architecture}

\begin{figure}[th!b]
\centering
\includegraphics[width=8cm, height=9.3cm]{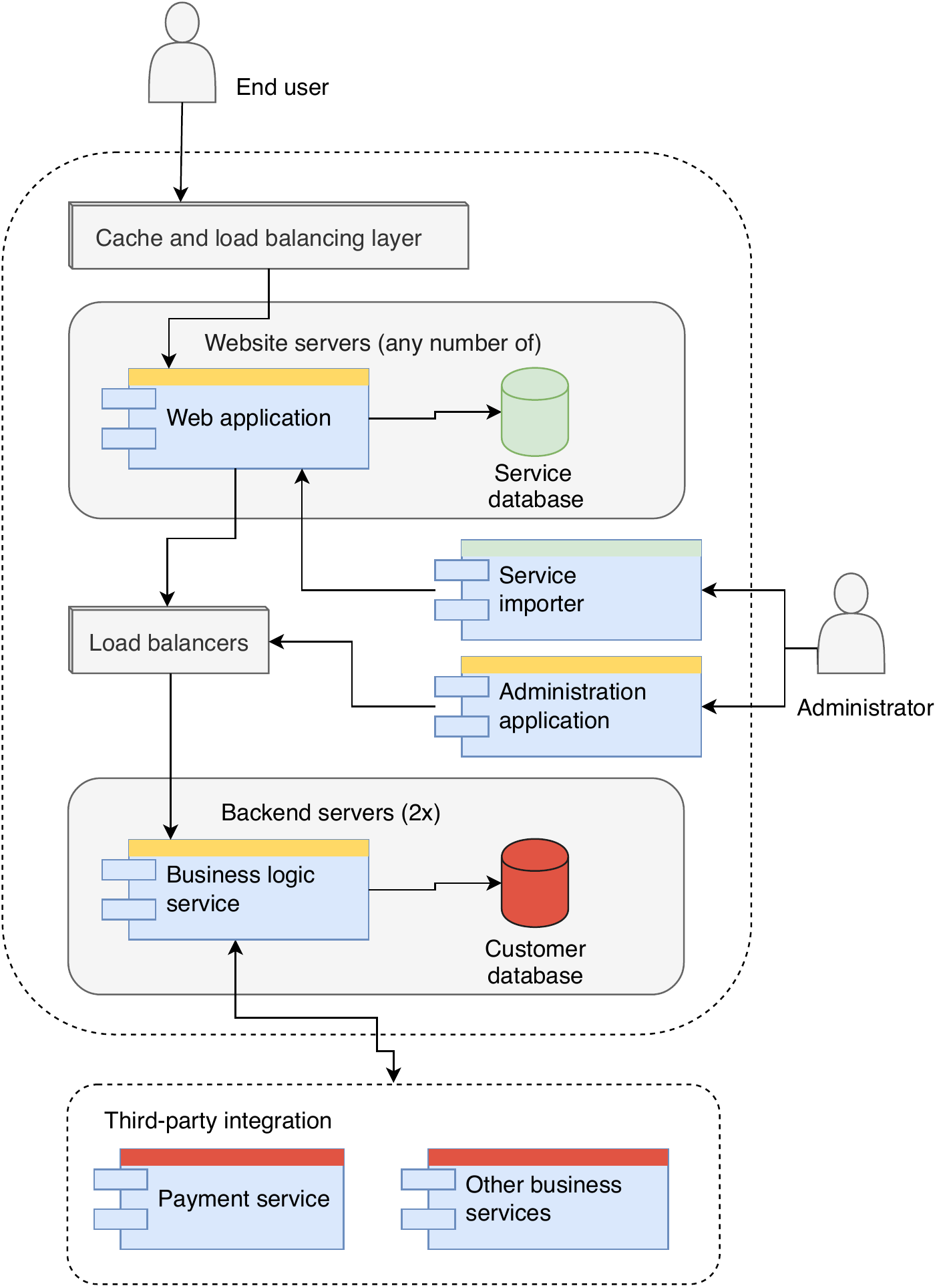}
\caption{An Example Architecture}
\label{fig: architecture}
\end{figure}

A brief concrete example can be used to further elaborate the constraints and the \change{company's} general operating environment. Thus, Figure~\ref{fig: architecture} displays \change{a high-level} example of a typical SOA within the case company. The colors used represent the sensitivity in the handling of \change{sensitive} personal data from a software architecture perspective: a red color corresponds with actual storage of such data, a yellow color illustrates the delivery of personal data, and elements marked with a green color are free from any handling of sensitive personal data. 

A user (consumer) of the case company's customer delivers personal data (such as contact details) to the architecture due to a subscription-based service offered by the customer. There are two databases to which such data is stored, but both of these are located within a single logical module. Given Constraint~$\constraintDistributionOfPersonalData$, the data distribution is therefore smaller compared to a solution in which personal data would be stored also to the database used at the presentation layer of the architecture. Given Constraint~$\constraintVarietyOfServers$, the storing of personal data is further isolated to separate servers. Given Constraint~$\constraintMaintenanceAndSupport$, also administration is separated. Although this example SOA is simple, fairly large changes are required due to the requirements elicited from the GDPR. These must operate within the contextual limits imposed by Constraints~$\constraintTechnologicalDependency$, $\constraintExternalInvestments$, $\constraintUpdateOfOldSystems$, and $\constraintReusability$.

\section{Requirements}\label{sec: requirements}

\subsection{\change{Articles, Architectures, and Constraints}}

The GDPR contains ninety-nine \textit{articles}~(A) arranged into eleven chapters. In what follows, a quick overview of the regulation is presented in order to justify the subsequent requirements elicitation. The brief discussion mainly pinpoints a few particularly relevant articles for software architectures (for more comprehensive legal discussion see \cite{Wachter18}, \cite{TikkinenPiri18}, and~\cite{Veale18}, for instance). \change{To provide a concise summary, these are discussed in conjunction with the constraints and software architecture design.} When referring to a particular recital in an article, the recital is placed in parenthesis. As an example: the~notation A32(1) refers to the first recital in Article~A32.

The initial four articles define the regulation's general provisions, material scope, territorial scope, and definitions. In essence, these state the obvious: the GDPR is about protecting natural persons and their personal data within the EU in the context of data processing by both automated and manual means. In terms of territorial scope, it should be remarked that the case company and its customers are operating within the EU, although Constraint~$\constraintCustomerHeterogenity$ implies that potential extraterritorial considerations apply at least in principle. Among other definitions and clarifications, Article~A4 defines the concepts of profiling and pseudonymization, which are both relevant \change{also} for the case company. While the details are beyond the scope of this paper, it can be remarked that pseudonymization was implemented in accordance with existing \change{suggestions~(see, in particular, \cite{Hintze18})}. Furthermore, A4(7) and A4(8) provide the definitions for data controllers and data processors. While these two concepts are difficult juridically~\cite{TikkinenPiri18}, the interpretation for the case company is clear: despite of the business model, the company is both a controller and a processor.

Articles A5 and A6 define the principles and lawfulness. Both are relevant for the case company's SOAs. In essence, particularly A5(1c), A5(1e), and A5(2) state that compliance can be achieved by demonstrating that only a minimal amount of relevant personal data is securely processed for a particular purpose, and the data's life cycle is well-controlled. Article A6 continues by noting that personal data can be lawfully processed only with a user's consent or in case the processing is backed by a contract, a legal obligation, or some related exemption. Article A7 further clarifies the conditions for a user's consent. While together A6(1a) and A7 imply a clear requirement for the case company, it should be underlined that the company's compliance is not relying on A6(1f) and its much debated~\cite{Wachter18} concept of so-called legitimate interests.

Articles from A12 to A23 define the rights of data subjects. To begin with, A12 sets the general principles for transparency and users' rights. While no technical details are enumerated, A12(7) and A12(8) remark about the desirability of using standardized icons for illustrating the intended processing of personal data. In theory, these may be useful also in terms of reusability ($\constraintReusability$). Articles A13 and A14 define the information to be delivered to users upon request. Particularly important are the clauses about the period of data storage and the ending of data processing. For the case company, these imply project-specific solutions, which are challenging due to the variety of customers ($\constraintCustomerHeterogenity$) and projects ($\constraintSizeOfProject$). Common frameworks and $\constraintReusability$ can balance these challenges. Because a user needs to generally make a request in order to exercise his or her rights, also the case company's support channels ($\constraintMaintenanceAndSupport$) are involved.

Articles A16, A17, A18, and A19 define the rights for rectification, erasure, and restriction. For software architectures these imply different technical considerations about acknowledgements, backups, and logs. In general, these considerations mandate that the distribution of personal data ($\constraintDistributionOfPersonalData$) within a software architecture has already been properly addressed.

Articles A20 and A21 define the rights for data portability and objection. In terms of the former, a noteworthy observation for software architectures is the emphasis on structured, widely used, and generally machine-readable formats. While no universally agreed format exists, a compliant architecture should still be prepared to deliver requested personal data in a well-defined format. A common interface is possible for A21.

Articles from A24 to A31 enumerate the obligations for data processors and data controllers. Article A25 is particularly important: it declares that rigorous data protection should be applied by default, albeit in the limits of state-of-the-art technical solutions (cf.~$\constraintTechnologicalDependency$) and implementation costs (cf.~$\constraintExternalInvestments$). Article A30 is also important for the design of software architectures: records should be kept from all processing activities involving personal data. This bookkeeping necessity is in line with the accountability criterion in A5(2). Regardless of a particular article or its recital, a fundamental prerequisite is that a software architecture is already well-manageable.

Finally, Articles A32 and A44 are worth pointing out. The former explicitly aligns data protection with information security. A conventional listing follows---from encryption and pseudonymization to CIA (\textit{confidentiality}, \textit{integrity}, and \textit{availability}), and from there to risks and likelihoods. While Articles A33 and A34 spell out the much discussed data breach notification mandates, these are less relevant for a software architecture design in the sense that notification functionality should be built-in~\cite{Kapitsaki18}. Article A44 together with the chapter five in the regulation define the conditions via which personal data of Europeans can be transmitted to outside of the EU.

\subsection{Elicitation of Requirements}

The preceding short discussion about the legal background allows to elicit nine general \textit{requirements} (R) for the case company's software architectures. These are enumerated in Figure~\ref{fig: requiremnets}. Many of the requirements listed are accompanied with their essential properties. Furthermore, Article A5(2) and its accountability criterion can be attached to each requirement. \change{A selected set of} minimal \textit{user stories} (U) is also \change{presented for illustrative purposes. Even though a comprehensive collection of stories was assembled for the actual implementation, the few cases presented are important for underlining that user stories are applicable also in the GDPR context. The point is worth stressing because user stories  are extensively used in agile software development~\cite{Cohn04}.} The nine requirements are:

\begin{itemize}
\item{\textit{System security and privacy} ($\requirementSystemSecurityAndPrivacy$) is the label used for the first requirement. It includes data protection (A25) and information security~(A32). Therefore, this requirement provides the fundamental basis for all other requirements elicited. If a verification of CIA indicates some major problems, the premises of all other requirements are obviously threatened. The requirement also faces the company's operational constraints. Due to Constraint~$\constraintCustomerHeterogenity$, some customers may have additional requirements for data protection, but the fundamental security aspects should apply to all projects. For instance, access control mechanisms must be implemented properly and security-related maintenance must be guaranteed~($\constraintMaintenanceAndSupport$). However,
it may also be that some particular customers require further security guarantees. Such additional requirements bring questions about technological choices~($\constraintTechnologicalDependency$), pricing, investments~($\constraintExternalInvestments$), and reusability ($\constraintReusability$). Finally, a large distribution of personal data ($\constraintDistributionOfPersonalData$) affects also security considerations because there is also a large attack surface.}
\item{\textit{Data minimization} ($\requirementDataMinimization$) is a rather clear requirement corresponding with A5(1c) and A25(1). Accordingly, the handling and storing of personal data should be minimized. Constraint~$\constraintDistributionOfPersonalData$ and more generally distributed software architectures cause some difficulties for meeting the requirement---it must be possible to show that all elements of a software architecture deal with personal data only minimally. What is more, it must be possible to demonstrate a minimal use of personal data in the staging environment ($\constraintVarietyOfServers$) used during development and testing.}
\item{\textit{Consent control} ($\requirementConstentControl$) refers to a requirement for handling personal data according to Articles~A6 and A7. This requirement is decomposed into the collection of users' consents ($\requirementConstentControl.1$), the abstract data structures used for these ($\requirementConstentControl.2$), and special considerations (A8) required for children and their guardians ($\requirementConstentControl.3$). The data structures include timestamps upon which consents were received, as well as a possibility to operate with digital signatures required for strong authentication.  In terms of constraints, reusability ($\constraintReusability$) and potential investments ($\constraintExternalInvestments$) are worthwhile to remark. Due to Constraint $\constraintCustomerHeterogenity$, consents must be collected also from users of old customers. Regarding user stories, a couple of simple examples could spell:
\begin{itemize}
\item{U3.1: \textit{As a user, I am able to give and cancel my consent for the processing of my personal data.}}
\item{U3.2: \textit{As a guardian of an underage user registered to a service, I am able to cancel my consent regarding the processing of my child's personal data.}}
\end{itemize}
}
\item{\textit{Data traceability} ($\requirementDataTraceability$) is a requirement stemming particularly from A30. Given the mandated ability to demonstrate compliance, traceability must include longitudinal considerations, users' requests, potential disclosures to third-parties, and potential transfers of personal data to outside of the EU. Careful logging is the obvious choice for addressing the requirement. For implementation, legacy systems ($\constraintUpdateOfOldSystems$), maintenance and support~($\constraintMaintenanceAndSupport$), and distribution of personal data (in terms of both $\constraintVarietyOfServers$~and~$\constraintDistributionOfPersonalData$) require attention. Reusable and flexible frameworks ($\constraintReusability$) engender some preferable synergies for implementation.}
\item{\textit{User access} ($\requirementUserAccess$) refers to A15 and A20: users must have access to their personal data in a sensible machine-readable format and they have a right for data portability. In general, both articles can be addressed with approximately the same amount of work; a user interface, access controls, and a well-defined data format must be implemented. Care should be taken to avoid accidentally divulging personal data of other users~\change{\cite{Veale18, Singh19}}. Like with $\requirementDataMinimization$, distributed systems, relational databases, and scattering of personal data ($\constraintVarietyOfServers$~and~$\constraintDistributionOfPersonalData$) bring further challenges. A~simple user story for $\requirementUserAccess$ might read like:
\begin{itemize}
\item{U5.1: \textit{As a user registered to a service, I am able to view all data stored about me in an understandable format, and transfer this data to another service.}}
\end{itemize}
}
\item{\textit{Data rectification} ($\requirementDataRectification$) covers A16. That is, users have a right to correct data stored about them, including data delivered to third-parties. In addition to the considerations and challenges listed with regards to $\requirementUserAccess$, it should be noted that backups introduce a special concern; data should be collected in all copies~\cite{Altorbaq17}. It must be also ensured that restoring data does not overwrite changes.}
\item{\textit{Data erasure} ($\requirementDataErasure$) stems from A17 and its famous ``right to be forgotten''. In addition to the points already discussed with respect to Requirements~$\requirementUserAccess$ and $\requirementDataRectification$, it should be also recalled that personal data must not be stored longer than necessary, as mandated by A5(1e). A~proper solution to $\requirementDataTraceability$ helps at achieving also this demand.}
\begin{figure}[t!]
\centering
\includegraphics[width=\linewidth, height=12.8cm]{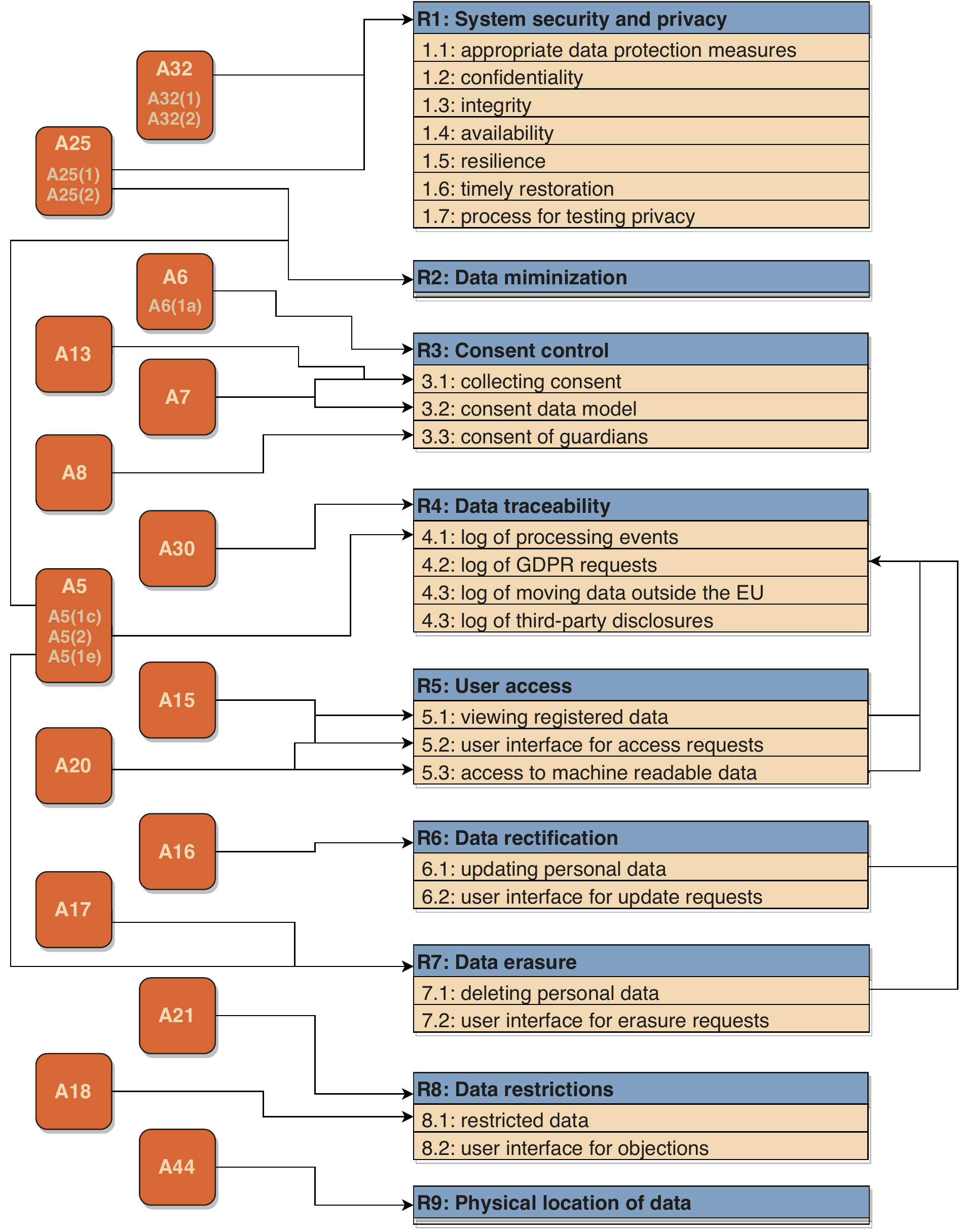}
\caption{Elicited Requirements from the GDPR}
\label{fig: requiremnets}
\end{figure}
\item{\textit{Data restrictions} ($\requirementDataRestrictions$) refers to A21 and its obligations for data controllers to allow data subjects to object the processing of their personal data, including but not limited to profiling. An objection again implies a request in practice, so a user interface is required together with access controls (R8.2). If the request is deemed as valid, the data associated can no longer be processed, modified, or deleted (A18). Thus, a particular logic is needed for handling restricted data (R8.1). In terms of implementation, this logic is analogous to the one used for consents~(R3). In fact,  both a consent and potential restrictions must be checked before processing personal~data of data subjects.}
\item{\textit{Physical location of data} ($\requirementPhysicalLocationOfData$) is the final requirement. Even though the case company is not doing business outside of the EU, it must be still known where personal data is located geographically. Knowledge is explicit in terms of the core software architecture, but~$\constraintTechnologicalDependency$ (technological dependency) may necessitate integration of different cloud services. Fortunately, all decent providers of these allow to specify the geographic location of servers. Constraints $\constraintVarietyOfServers$ and $\constraintMaintenanceAndSupport$ also breed some implicit concerns. For instance, personal data may accidentally spill to outside of the EU due to as innocent reasons as the uses of version control and documentation systems during software development. Given the low privacy awareness among software developers in general~\cite{Hadar18}, education and instructions are the generic means to address these concerns. A data protection officer mandated by A37 can help at this task. Another good option is to write a detailed architectural description akin to this~paper.}
\end{itemize}

\section{\change{Architectural Changes}}\label{sec: architecture}

\begin{figure}[th!b]
\centering
\includegraphics[width=8cm, height=9.3cm]{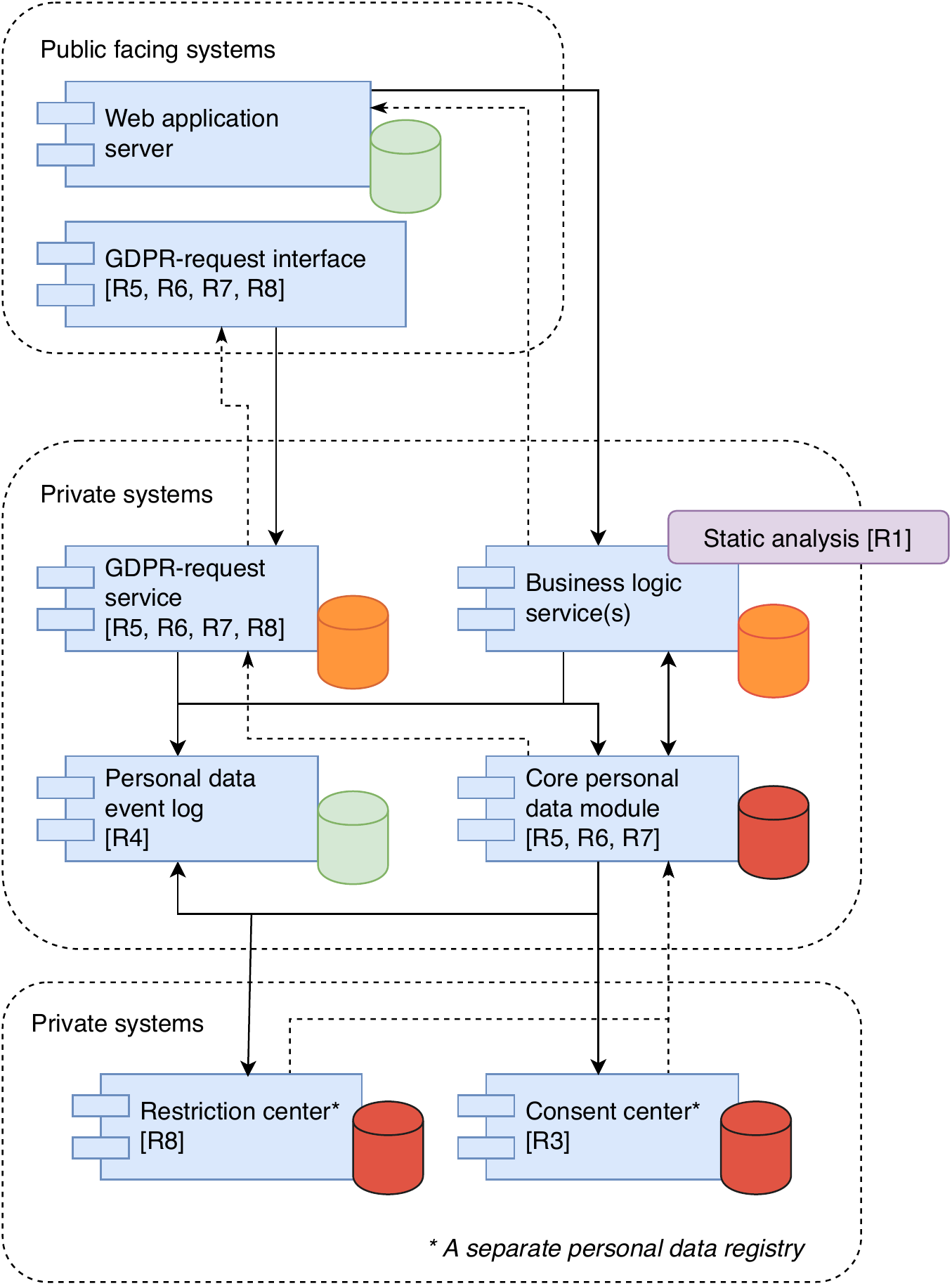}
\caption{An Overview of the Architectural Changes}
\label{fig: architectural overview}
\end{figure}

The nine requirements elicited from the GDPR provide the basis for the changes required to the SOAs used in the case company. In what follows, the principal architectural changes and choices are thus elaborated. Requirements $\requirementSystemSecurityAndPrivacy$ and $\requirementPhysicalLocationOfData$ are excluded from the elaboration. The former sets the general security and privacy requirements that span the whole architecture---and beyond, whereas the latter is a unique requirement that cannot be explicitly addressed by the means of architectural design. Given these remarks, Figure~\ref{fig: architectural overview} shows an overview of the redesign changes induced by the requirements. 

\begin{figure}[th!b]
\centering
\includegraphics[width=7.5cm, height=6.1cm]{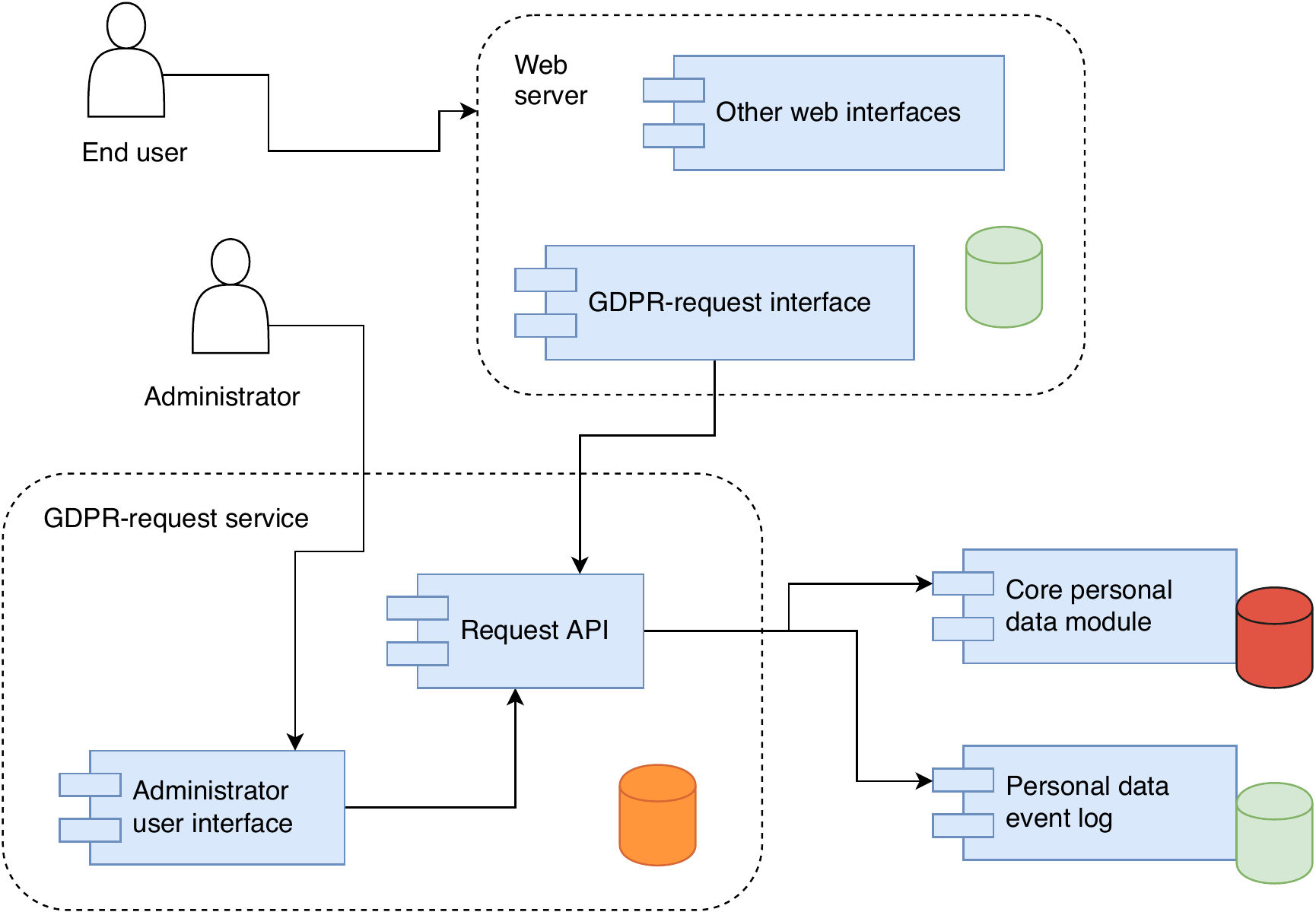}
\caption{A Design for GDPR Requests}
\label{fig: request design}
\end{figure}

There are four requirements that imply a user interface through which data subjects can exercise their rights. While these can be dressed with project-specific themes and styles, the logic is similar for all instances. Therefore, a~common GDPR-request interface is used for $\requirementUserAccess$, $\requirementDataRectification$, $\requirementDataErasure$, and~$\requirementDataRestrictions$. The interface should not be interpreted as the only possible way for users to exercise their rights: if necessary, requests can be processed also manually based on inquiries received through other channels. From a software architecture perspective, the logic of processing the requests should be the same either way.

Because users are registered to the services offered via subscriptions, basic access control mechanisms apply. Due to abuse risks and potential legal requirements, these are not enough, however. The GDPR-request interface is therefore accompanied with a specific GDPR-request service accessed by administrators through a light user interface of their own. In other words, an administrator approves or declines a GDPR-specific request made by a user already authenticated to the public facing systems. The mapping between these two interfaces is further illustrated in Figure~\ref{fig: request design}. In-between the two interfaces is a specific \textit{application programming interface} (API) for handling the request-response dynamics. In theory, automation might be possible through this API, but, in practice, the dynamics are delayed; a user makes a request, which is later handled by a given service-specific administrator (that is, a support person). The only exception is Requirement~$\requirementUserAccess$ for which also automation might be considered. In any case, delayed requests require storing these to persistent storage. Because of the data traceability requirement~($\requirementDataTraceability$), storage is also explicitly required regardless of the handling dynamics.

\begin{figure*}[t!]
\centering
\includegraphics[width=14cm, height=6.99cm]{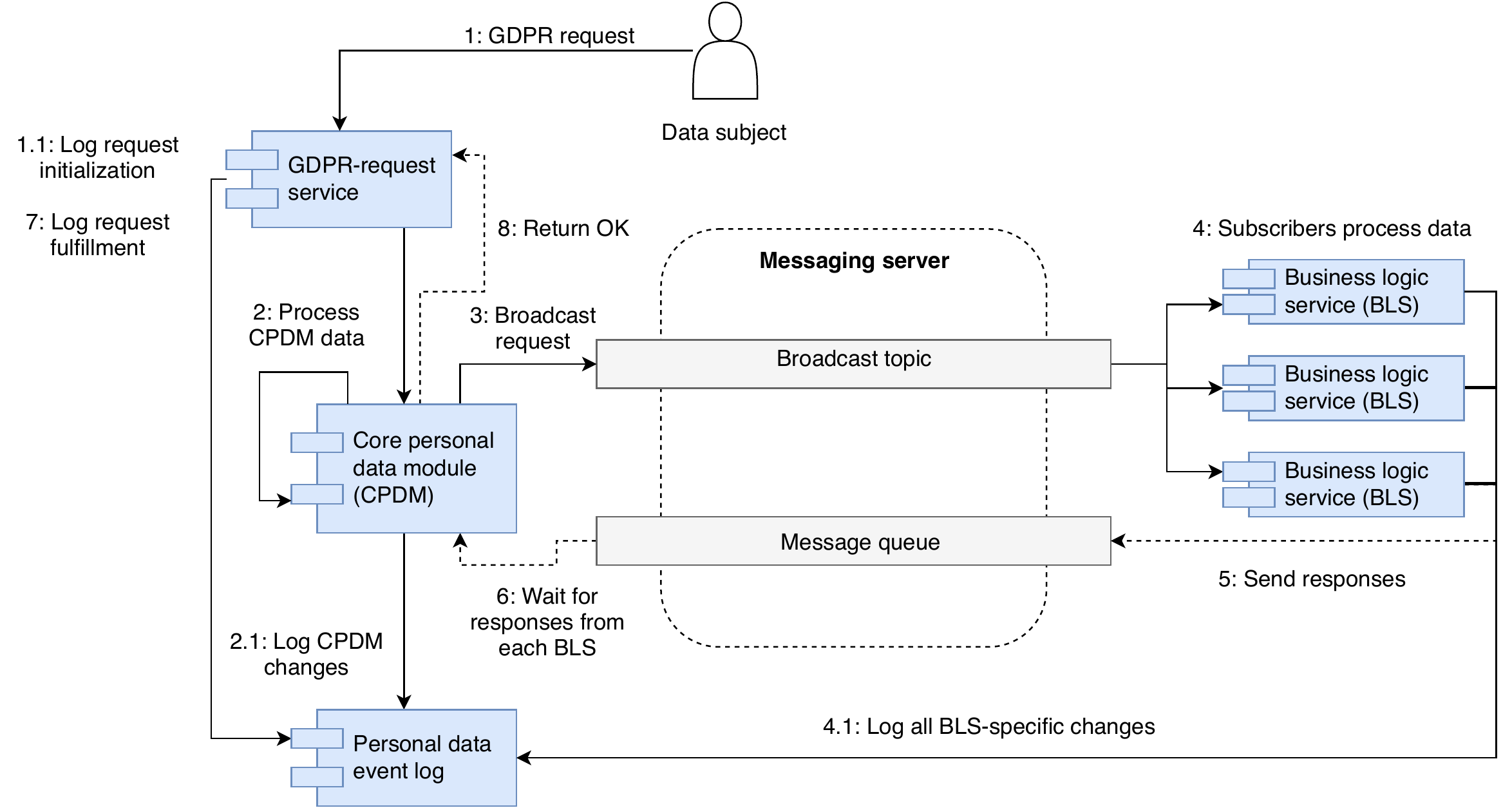}
\caption{A Messaging Service Bus}
\label{fig: messaging}
\end{figure*}

As earlier (see Subsection~\ref{subsec: an example architecture}), the colors of the databases in Figures~\ref{fig: architectural overview} and \ref{fig: request design} connote with the sensitivity of personal data. As can be seen, the request-specific databases are marked with an orange color; some personal data must be stored to these databases for adequately handling the requests, but the scope of this data is limited. The varying scope or depth of the personal data leads to an important architectural design choice: the handling and storage of all truly sensitive personal data is isolated into a specific \textit{core personal data module} (CPDM). For instance, the databases holding information about GDPR-requests may contain user names and email addresses, whereas those within the CPDM may hold users' real names or, in theory, even their social security numbers. While this distinction between ``regular'' and sensitive data is commonly made~\cite{Kapitsaki18}, it should be further noted that the distinction also follows the different theoretical levels of pseudonymization~\cite{Hintze18}. In other words, the CPDM holds all information from which a user can be directly identified. Although $\requirementDataMinimization$ is difficult to address in general, this distinction provides the architectural basis for addressing the GDPR's data minimization requirement. The centralization also helps in meeting the general design principle about reducing the distribution of personal data ($\constraintDistributionOfPersonalData$).

Because of the data restriction requirement ($\requirementDataRestrictions$), operations (additions, deletions, or updates) in the CPDM's database(s) need to verify whether users have placed restrictions for these. An analogous logic applies to the consents ($\requirementConstentControl$). Thus, further isolation is used for~$\requirementConstentControl$ (consent center) and~$\requirementDataRestrictions$ (restriction center). The CPDM consults these two centers before accessing or altering its underlying database(s). The implementation of these follows closely the ideas behind the so-called MyData framework~\cite{SuHyysalo16}. Therefore, the centers may be easily altered to use also consent services provided by external operators.

As for the collection and potential cancellation of consents, these can occur either within the architecture itself or through services implemented in the specific consent center. The latter services can be implemented in accordance with well-established authentication and authorization protocols. The notable examples in this regard are OAuth and~OpenID. As the data structures about consents ($\requirementConstentControl.2$) holds also the dates upon which these were received from users, it is possible to also address the GDPR's general mandate about fixed retention periods~(A5). For instance, it is possible to periodically scan the database of the consent center to ensure that the architecture does not commit ``the sin of storing data forever''~(cf.~\cite{Shastri19}). If such automatic scanning indicates particularly old records, requests about consents can be resend, or, in some cases, the personal data can be outright removed \change{in an automated manner.}

In terms of the overall software architecture, it is important to continue by noting that each \textit{business logic service} (BLS) needs to have a reference to the sensitive data stored to the CPDM. These services are internal to the case company and do not refer to the services of the company's customers. For instance, in Fig.~\ref{fig: architecture} a BLS handles an integration with a third-party payment service. For this reason, the CPDM and the business logic services can be reasonably also placed into the same logical module (see Fig.~\ref{fig: architectural overview}). To ensure data integrity ($\requirementSystemSecurityAndPrivacy$.3) and traceability ($\requirementDataTraceability$), changes made through the GDPR-request interface must propagate to the databases of each BLS. To ensure that $\requirementDataErasure$ is properly met, all data must be deleted, regardless whether this data is sensitive or not. Likewise, the CPDM must be able to collect the data  from each BLS in order to comply with $\requirementUserAccess$ and $\requirementDataRectification$. The same applies with respect to~$\requirementDataRestrictions$: if a user has restricted the use of his or her data in order to pursue or defend legal claims (A18), also the data held by each BLS must be frozen. The required functionality is implemented in the architecture by using a message bus through which the CPDM broadcasts the requests to each BLS that has registered to the bus, and, therefore, declared using non-sensitive but still personal data of users. By excluding the consent and restriction centers for the sake of clarity, the logic of the broadcasting functionality is illustrated in Figure~\ref{fig: messaging}.

The final architectural design choice is about the traceability of personal data ($\requirementDataTraceability$) and the logging required to implement this requirement. In essence, the guiding question is: what should be logged and how? To begin with the more important what-question, the obvious starting point is that no sensitive personal data whatsoever ends up in the logs. By asserting that this necessary condition holds, there are four types of events that should be logged according to the requirements elicitation. 

First, the processing of personal data should be logged with sufficient detail ($\requirementDataTraceability.1$). In the case company's software architecture such logging implies verifying that each BLS is implementing logging with sufficient rigor. Particular attention is required to ensure that logging is systematically and correctly implemented in those cases involving automated decision-making and profiling~(A22). Second, all events about GDPR requests should be logged ($\requirementDataTraceability.2$). As can be seen from Figure~\ref{fig: messaging}, this type of logging is rigorously implemented in the CPDM module. Because the GDPR mentions in several occasions that undue delays must be avoided, this type of logging is necessary to guarantee the legal rights of both data subjects and the case company.  Given the message bus, this logging not only follows the general ``request-log-process-log'' logic often used in the data protection (privacy) context~\cite{Antignac18}, but also accounts for the activities of each BLS. Third and fourth, all disclosures to third-parties are logged alongside the supply of data to outside of the European Union. As said, no data is presently transmitted outside of the union, and a disclosure to third-parties is a rare event. Therefore, the logging for $\requirementDataTraceability.3$ and $\requirementDataTraceability.4$ is mainly implemented to ensure an audit trail for potential accidents involving data spillage.

\begin{figure}[th!b]
\centering
\includegraphics[width=8cm, height=4.3cm]{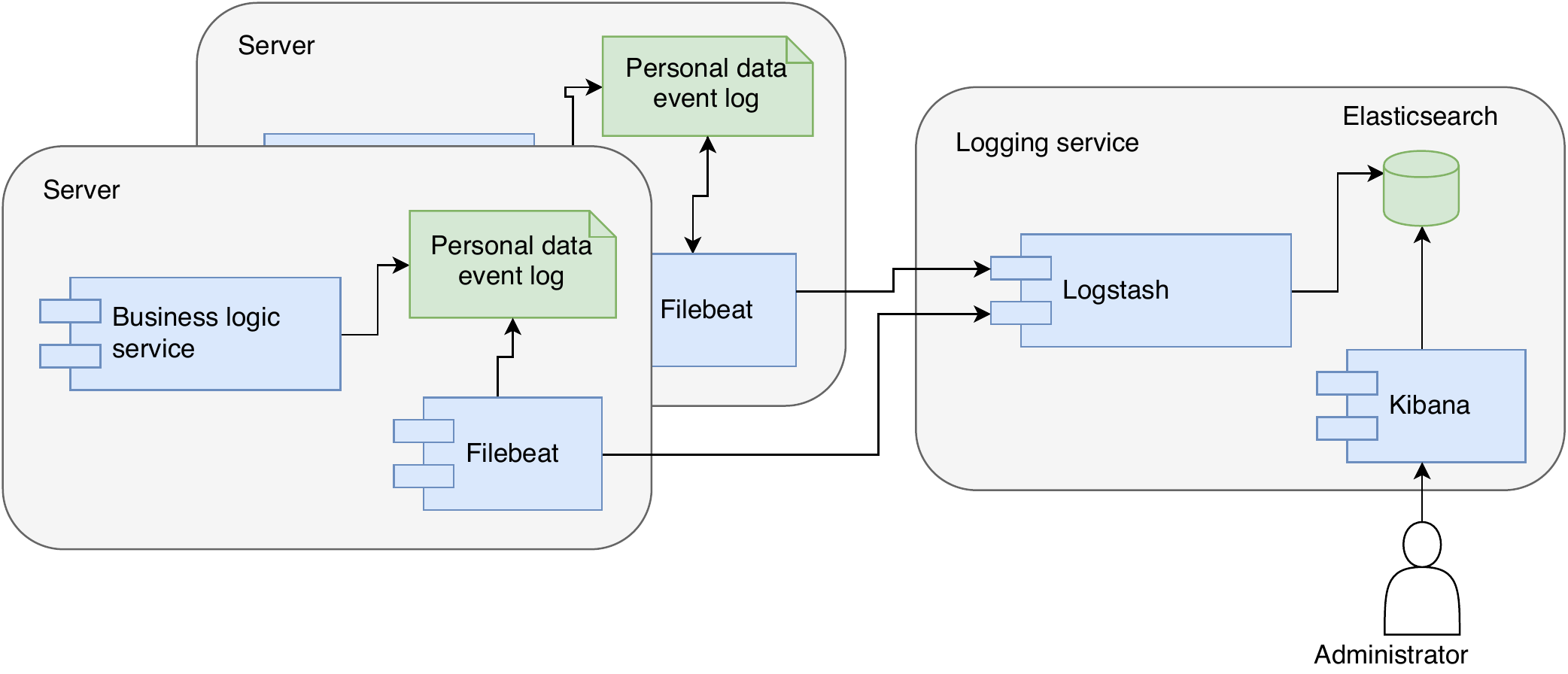}
\caption{Logging Architecture}
\label{fig: logging}
\end{figure}

As for the how-question, centralization is again a sensible design goal: administrators should be able to access all logs through a single interface. Because no personal data is stored to the logs, a so-called ELK-stack (Elastic, Logstash, and Kibana) offers a plausible implementation choice. This stack is illustrated in Figure~\ref{fig: logging}. In essence: an administrator uses a visualization tool (Kibana) for analyzing the logs based on a text-based search engine (Elasticsearch), which relies on an input stream (Logstash) for collecting the logs from each server given a service (Filebeat) installed to these servers.

Finally, it is worthwhile to briefly return to Requirement~$\requirementSystemSecurityAndPrivacy$. While the requirement's information security side---including the fundamental CIA-triad---is outside of the scope of this paper, the data protection side still warrants a remark. To improve the prerequisites for testing (R1.7), as well as other related imperatives~\cite{Altorbaq17}, the whole software architecture was audited by annotating all locations and services that handle personal data. All of these can be thus located automatically through static analysis given the annotations \textit{@PersonalData} and \textit{@PersonalDataHandler} used during the auditing process.

\section{Discussion}\label{sec: discussion}

\subsection{Questions Answered}

\begin{figure}[th!b]
\centering
\includegraphics[width=\linewidth, height=15.4cm]{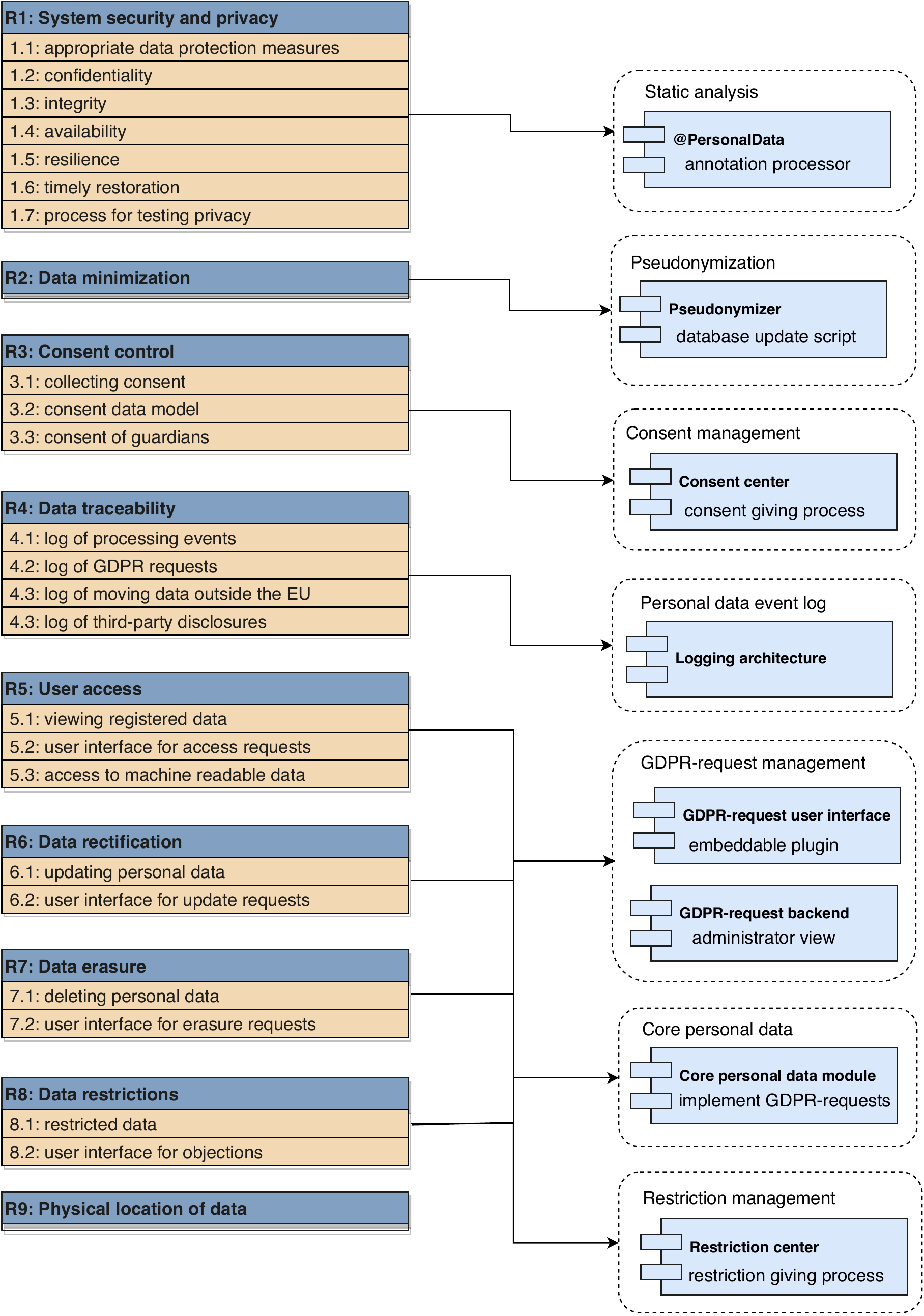}
\caption{A Summary of the Requirements and Design Choices}
\label{fig: summary}
\end{figure}

This paper used a grounded theory approach for addressing the GDPR's main implications for software development SMEs. The approach was framed around constraints, requirements, and software architectures. The answers to the corresponding research questions can be summarized as~follows:

\begin{itemize}
\item{\researchQuestionConstraints: There are many practical constraints that typical software development SMEs face when implementing changes required by the GPDR \change{to SOAs}. Given the case company's business model and operating environment, the paper identified nine practical constraints. These can be roughly grounded into three categories: \textit{business} $(\constraintCustomerHeterogenity,
\constraintTechnologicalDependency,
\constraintSizeOfProject, \constraintExternalInvestments )$, \textit{infrastructure} $( \constraintVarietyOfServers, \constraintDistributionOfPersonalData)$, and \textit{software engineering} $( \constraintUpdateOfOldSystems, \constraintReusability, \constraintMaintenanceAndSupport)$ constraints. All three groups are important for requirements engineering. While the business constraints have stolen much of the attention in media, different infrastructural and software engineering constraints are arguably more important when implementing various changes induced by the GDPR in practice.}
\item{\researchQuestionRequirements: Despite of the GDPR's immense length and scope, only nine relevant requirements were elicited for the case company (see Fig.~\ref{fig: requiremnets}). While many of these include their specific subcategories, the bottom line is that the GDPR's demands \change{for software architectures} are addressable with fairly generic requirements \change{at least in the SME-SOA context studied}. The requirements elicited can be grouped into five categories: \textit{privacy and security} $(\requirementSystemSecurityAndPrivacy)$, \textit{data minimization}~$(\requirementDataMinimization)$, users' \textit{rights} ($\requirementConstentControl, \requirementUserAccess, \requirementDataRectification, \requirementDataErasure, \requirementDataRestrictions)$, \textit{traceability}~$(\requirementDataTraceability$), and physical \textit{location} of personal data~$(\requirementPhysicalLocationOfData)$.}
\item{\researchQuestionArchitectures: Given the constraints identified, the requirements elicited were addressable with relatively simple changes to the already existing software architectures. The main technical solutions are further summarized in Figure~\ref{fig: summary}. In general, \textit{user interfaces}, \textit{isolation} of personal data, \textit{access control} mechanisms, \textit{pseudonymization}, \textit{logging}, and \textit{annotations} were the biggest implementation changes required. Although the changes themselves are specific to the case company, the architectural design choices \change{may} generalize also to other software development~SMEs.}
\end{itemize}

\subsection{\change{Limitations and Challenges}}

\change{A notable limitation relates to generalizability. The paper's explicit framing to SMEs obviously limits generalizability. The further framing to SOAs limits it further. The GDPR may place rather different requirements for a SME operating in the Internet-of-things domain, for instance. That said, the generalizability problems should not be exaggerated. As the MyData framework has demonstrated~\cite{SuHyysalo16}, general design patterns (such as the ones in Fig.~\ref{fig: architectural overview} and Fig.~\ref{fig: messaging}) are applicable in many different contexts and industry sectors. Another point about generalizability relates to the fact that only those GDPR articles were considered that have clear implications for software architectures. While this framing is common in the requirements engineering domain~\cite{ChenBabar13}, it conceals the GDPR's many non-technical requirements that are important also for SMEs~\cite{Brodin19}. To this end, concepts such as functional and non-functional requirements could be used to systematically analyze and categorize the individual articles and their recitals.}

The argumentation-solution layering discussed in Section~\ref{sec: background} provides also a good way to discuss a few practical challenges and problems faced. Validation of the requirements elicited is the primary issue at the argumentative layer. Besides conforming with the GDPR itself, the nine requirements elicited should generally be consistent, unambiguous, implementable, verifiable, and traceable~\text{\cite{AyalaRivera18, Bobkowska10}}. Because concrete solutions were implemented, most of the requirements satisfy these validation attributes. The only exception is \change{$\requirementSystemSecurityAndPrivacy$. It} is seldom---if ever---possible to write down a requirement (set) that would be specific enough to capture all possible threats and risks. As has been suggested~\cite{Elluri18}, a possible path forward might be to combine the GDPR's requirements with those from other standards, such as the Payment Card Industry Data Security Standard~(PCI~DSS). \change{More generally, security-specific standards and governmental guidelines (such as the Finnish VAHTI) should arguably be augmented with material on data protection and privacy. The point applies also to software and requirements engineering standards, such as the ISO/IEC/IEEE 29148:2018 for software life cycle management.}

Four noteworthy challenges can be mentioned at the solution layer. First, it remains somewhat unclear whether the pseudonymization solution is sufficient to ensure the ``appropriate safeguards'' mentioned in several occasions in the GDPR. While all personal data is further encrypted and everything is transmitted through the transport layer security protocol, the practical challenge rather relates to the logical fact that personal data is not pseudonymized at the level of the whole architecture due to the references between the CPDM and the business logic services. Pseudonymization can be guaranteed with respect to each BLS, however. Given the company's business model, this guarantee seems sufficient.

Second, it has been argued that annotation and static analysis are not enough because these miss the purpose of using personal data~\cite{Basin18}. While the pseudonymization solution partially addresses also this concern, it must be acknowledged that the intention of using personal data is not made explicit in the architecture. A possible solution might be an additional authorization module that each BLS would be mandated to use prior to registering to the broadcasting service bus in Figure~\ref{fig: messaging}.

Third, distributed software architectures seem a double-edged sword when it comes to data protection and privacy. While these provide great opportunities for modularity and server isolation~\cite{Kittman18}, these also increase the risk that personal data accidentally spills to unintended servers, modules, third-parties, or geographic locations. The last problem~follows.

Fourth, requirements should be traceable through a whole software development life cycle~\cite{AyalaRivera18}, and even through a whole software life cycle, from the first working release to the eventual deprecation. Insofar as practical software engineering is considered, this traceability mandate is particularly interesting. As was noted in Subsection~\ref{subsec: constraints}, a particular concern in the present work relates to the staging environment used during development and testing, but the problem goes beyond such environments. For instance, the logging architecture in Figure~\ref{fig: logging} rests on the assertion that no sensitive personal data is logged. While the assertion can be assured to hold in the case company, the example is illustrative in the sense that also various software development practices, tools, and frameworks have direct consequences for complying with the GDPR.

\subsection{Further Work}

\change{In addition to addressing the limitations and practical challenges discussed, a} couple of prolific paths can be noted for further work. \change{The first relates to the notions such as ``privacy-by-design'' and their practical software engineering equivalents.} The literature on software development and software architectures has long enjoyed a tradition of excellent textbooks and reference guidebooks. The examples include various monographs on design patterns, code smells, and related topics. Yet, very little has been written about design patterns for privacy and data protection, or about ``privacy smells'' and related topics. \change{The second path involves the question on how privacy and data protection requirements and their implementations could be systematically tested. Although static analysis was used for meeting $\requirementSystemSecurityAndPrivacy$.7, further research is required for developing new innovative testing and auditing techniques. These are necessary for reaching the GDPR's goal~(A42) about certification for privacy and data protection.}

\pagebreak
\clearpage
\bibliographystyle{IEEEtran}


\end{document}